\documentclass[twocolumn,superscriptaddress,showpacs,aps]{revtex4}
\usepackage{makeidx}
\usepackage{bm}
\usepackage[dvips]{epsfig}
\usepackage{graphicx,color}
\newcounter{fig}

\begin{document}
\title{Anisotropy of  graphite optical conductivity}
\author{ L.A. Falkovsky}
\affiliation{L.D. Landau Institute for Theoretical Physics, Moscow 119334, Russia }
\affiliation{Laboratoire de Physique des Solides, Univ. Paris-Sud, CNRS, UMR 8502, F-91405 Orsay Cedex, France }


\date{\today}      

\begin{abstract}
The graphite conductivity is evaluated for frequencies between 
 0.1 eV,
the energy of the order of the electron-hole overlap, and 1.5 eV, the electron nearest hopping energy.
The  in-plane conductivity per single atomic sheet is close to the universal graphene conductivity  $e^2/4\hbar$ and, however, contains a singularity conditioned by  peculiarities of the electron dispersion. The conductivity is less in the $c-$direction by the factor of the order of 0.01  governed by  electron hopping in this direction.

\end{abstract}
\maketitle

Recently, the light transmittance of graphene was found \cite{Na,Li,Ma} in the wide frequency region to  differ from unity by the value of  $\pi\alpha$, where  $\alpha$ is the fine structure constant   of  quantum electrodynamics.
These experimental observations are in  excellent agreement with the theoretical calculations  \cite{GSC,FV} of the graphene conductance, 
$G=e^2/4\hbar$, which does not depend on any material parameters.

This phenomenon is remarkable in two aspects. First, the fine structure constant has been found in one measurement for the first time in solid state physics. Second and most important, the Coulomb interaction does not disturb the agreement
between the experiment and the theory \cite{Mi,SS}. It should be emphasize that the Coulomb interaction in graphene is  poorly screened while the carriers are absent in this gapless insulator. 

In connection with this, it is interesting to study the change in the optical conductivity going from 2d graphene to its close "relative"\,, 3d graphite, with the optical conductivity  measured in Refs. \cite{TP,KHC}.

The electron properties of graphite is well described within the classical  Walles-Slonczewski-Weiss-McClure  theory \cite{SW}.
There are many parameters in this theory of the various order of value (see, e. g. \cite{PP}).   
Among them, the energy $\gamma_0= 3.1$ eV is largest  one representing the electron in-plane hopping between  nearest neighbors at the distance $r_0=$1.42 $ \AA$. If we are interested in  frequencies less than 3.1 eV, we can use the power $\bf{k}-$momentum expansion of the corresponding term in the Hamiltonian, taking only the linear approximation. Then the constant velocity $v=10^8$ cm/s  appears. The parameter $\gamma_1\simeq 0.4$ eV known from  optical studies of bilayer graphene \cite {KCM,Ba} is  next in the order. It describes the  interaction  between the nearest layers at the distance $c_0$=3.35 $\AA$.
   The parameters $\gamma_3$ and $\gamma_4$ give  corrections of the order of
10\% to the  velocity  $v$. Finally, the parameters $\gamma_2$, $\gamma_5$  of the order of 0.02 eV from the third sphere 
are used in order to describe the dispersion of the conduction and valence bands in the   $c-$direction. They are usually included in order to characterize the carriers and are known from the Shubnikov-de Haas oscillations and the cyclotron resonance.
 However, for the optical transitions at  relative high frequencies  $\gamma_2,\gamma_5\ll\omega\ll\gamma_0$, we can, first,
  neglect the smallest parameters $\gamma_2,\gamma_5$ and, second, use the linear ${\bf k}-$expansion with the constant velocity $v$ for in-layer directions. Our results have
the explicit analytic form.
\begin{figure}[]
\resizebox{.25\textwidth}{!}{\includegraphics{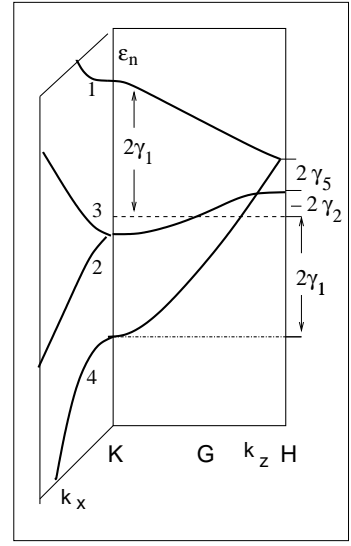}}
\caption{Dispersion of  low energy bands in graphite. } \label{gr}
\end{figure}

Thus, the simplified Hamiltonian of the model is given by
\begin{equation}
H(\mathbf{k})=\left(
\begin{array}{cccc}
0 \,    & k_{+} \,& \gamma(z)    \, & 0\\
k_{-} \,& 0     \, & 0\,& 0\\
\gamma(z)    \,  &0 \,& 0 \,  &k_{-}\\
0 \,& 0 \,&k_{+} \,&0
\end{array}%
\right) ,  \label{ham}
\end{equation}%
where the velocity parameter $v $  is included in the definition of the momentum components 
$k_{\pm}=v(\mp ik_x-k_y)$, and the constant $\gamma_1$ stands in the function 
$\gamma(z)=2\gamma_1\cos{z}$ depending on the dimensionless $k_z-$component  $z= k_zc_0$ along the $c$-axis,  $0<z<
\pi/2$.

The eigenenergies of the Hamiltonian write: 
\begin{eqnarray}\label{banddis}\varepsilon_{1,2}=\frac{\gamma(z)}{2}\pm\sqrt{\frac{\gamma^2(z)}{4}
+k^2}\,,\\ \nonumber 
\varepsilon_{3,4}=-\frac{\gamma(z)}{2}\pm\sqrt{\frac{\gamma^2(z)}{4}
+k^2}\,. 
\end{eqnarray}
The so-called "Dirac" point of graphene, $k=0$, turns into the K-G-H line  of the  graphite Brillouin zone, where the valence and conduction bands slick together, $\varepsilon_{2,3}=0$. It should be emphasized that this degeneration is conditioned by the lattice symmetry but is not  a result of the model.

Others two bands, $\varepsilon_{1,4}=\pm \gamma(z)$, are spaced at the distance  $ \gamma(z)$ which  vanishes  at the H point of the Brillouin zone. This band schema corresponds to the gapless semiconductor. 

In order to calculate the optical conductivity, we use the general expression
\cite{FV}
\begin{eqnarray}  \label{con}
&&\sigma ^{ij}(\omega) = \frac {2ie^{2}}{(2\pi)^3}\int
d^3k\sum_{n\ge m}\left\{ -\frac{df}
{d\varepsilon_n}\frac{v_{nn}^{i}v_{nn}^{j}}{\omega+i\nu}\right.
\nonumber \\&&+2\omega
\left. \frac{  v_{nm}^{i}v_{mn}^{j}[f(\varepsilon
_{n})-f(\varepsilon _{m})]}{%
(\varepsilon _{m}-\varepsilon
_{n})[(\omega+i\nu)
^{2}-(\varepsilon _{n}-\varepsilon _{m})^{2}]}\right\}\,,  
\end{eqnarray}
valid in the collisionless limit
 $\omega\gg\nu$, where $\nu$ is the collision rate. This condition is definitely fulfilled, if the frequencies are larger than the electron-hole overlap in graphite determined by the parameters $\gamma_2,\gamma_5$. The temperature is involved here by the Fermi-Dirac function
$f(\varepsilon)=[\exp(\frac{\varepsilon-\mu}{T})+1]^{-1}$,  the coefficient 2 takes into account the spin degeneration, and the integral is taken over the Brillouin zone where the electron dispersions $\varepsilon_n$ are defined. 

The first term in Eq.  (\ref{con}) is the intraband Drude-Boltzmann conductivity with the group velocity
$$\mathbf{v}_{nn}=\partial \varepsilon_n/\partial {\bf k}.$$
 This conductivity behaves as  $1/\omega$ and becomes less than the second term for frequencies higher than the electron-hole overlap. The second
 term corresponds with the  electronic interband transitions accompanied by the photon absorption. It involves the matrix elements of the velocity operator 
 \[U^{-1}\frac{\partial H({\bf k})}{\partial {\bf k}}U,\]  	
calculated in the representation transforming the Hamiltonian (\ref{ham}) to the diagonal form with the help of the operator $U$. We find for various transitions
\[\begin{array}{c}
v_{23}^x=2i(\varepsilon_3-\varepsilon_2)k_y/N_2N_3\,,\\
v_{12}^x=2(\varepsilon_1+\varepsilon_2)k_x/N_1N_2\,,\\
v_{14}^x=2i(\varepsilon_4-\varepsilon_1)k_y/N_1N_4\,,\\
\end{array}\]
where \(N_{n}^2=2(\varepsilon_n^2+k^2)\)\,.

The calculations show that  the off-diagonal components of conductivity reduce to zero and the in-plane diagonal  components are equal. For their real part, we obtain the integral which is explicitly taken over $\varphi$ and $ k$ in the polar coordinates at the zero temperature. Thus, we meet the integral over  $k_z$:
\begin{eqnarray}\label{resig}
\text{Re}~\frac{\sigma}{\sigma_0}=\frac{1}{\pi}\int_{0}^{\pi/2}dz\left[\frac{2\gamma(z)+\omega}{\gamma(z)+\omega}\right.\\ \left.
+\frac{2\gamma^2(z)}{\omega^2}\theta_1
+\frac{2\gamma(z)-\omega}{\gamma(z)-\omega}\theta_2\right]\,,
\nonumber\end{eqnarray}
where $\gamma(z)=2\gamma_1\cos{z}$, and $\theta_1$, $\theta_2$ are the step functions depending on  $\omega-\gamma(z)$ and $\omega-2\gamma(z)$, respectively. This integral can also be taken, but the result looks more complicated.

\begin{figure}[h]
\resizebox{.5\textwidth}{!}{\includegraphics{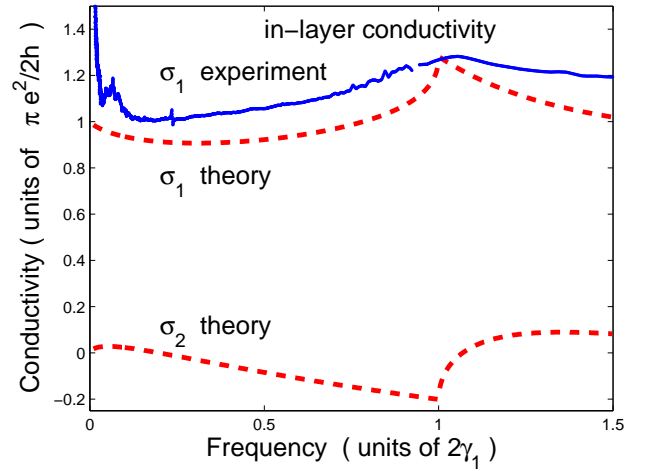}}
\caption{Real $\sigma_1$ and imaginary $\sigma_2$ parts of the graphite optical conductivity for the in-plane direction (per one atomic sheet in units of $e^2/4\hbar$) versus frequency (in units of $2\gamma_1=0.84$ eV); experimental data \cite{KHC},  solid line; results of the present theory, dashed lines.   } \label{gr1}
\end{figure}

Here, we introduce the conductivity
$\sigma_0=e^2/4\hbar c_0$ which can be named as the graphite universal  conductivity. It differs from the graphene conductivity only in the factor  $1/c_0$ which is simply the number of the atomic sheets in graphite per  length unit in the $c-$direction.  One can see, that the graphite conductivity goes to  $\sigma_0$ at low as well as high frequencies compared to  $\gamma(z)$
(see, also Fig. \ref{gr1}).  However, at $\omega=2\gamma_1$=0.84 eV, both
the kink   and the threshold are seen in the  real  and imaginary parts, correspondingly.  These singularities arise due to the electron transitions between bands  $2\rightarrow 1$ and $4 \rightarrow 3$ (see, Fig. \ref{gr}) described by the second term in Eq. (\ref{resig}).  The position of the singularities gives the value  $\gamma_1$=0.42 eV, which agrees well with  optical studies of bilayer graphene. 

Let us consider next the conductivity in the $c-$axis.  We need now the matrix elements $v_{nm}^z$. Calculations show that they are nonzero only for the transitions $2\rightarrow 1$ and $4\rightarrow 3$:
 \[ v_{21}^z=2\gamma^{\prime}(z)\varepsilon_1\varepsilon_2/N_1N_2\,,\]
\[ v_{43}^z=-2\gamma^{\prime}(z)\varepsilon_3\varepsilon_4/N_3N_4\,,\]
where the derivative $\gamma^{\prime}(z)=2\gamma_1c_0\sin{z}$.
Using Eq. (\ref{banddis}), we get
\[v_{21}^z= -v_{43}^z=-\gamma^{\prime}(z)k/\sqrt{\gamma^2(z)+4k^2}.\]
Integrating in Eq. (\ref{con})  over $\varphi$  and $ k$, we obtain 
\begin{eqnarray}\nonumber\text{Re}~\frac{\sigma^{zz}}{\sigma_0}=\left(\frac{\gamma_1c_0}{\hbar v}\right)^2I(t)\,,
\label{condz}\end{eqnarray}
where the integral over $k_z$ 
\[I(t)=\frac{4}{\pi}\int_0^{\pi/2}dz\sin^2{z}\left(1-\frac{\cos^2{z}}{t^2}\right)\theta(t-\cos{z})\]
with $t=\omega/2\gamma_1$.
This integral can be also taken and it has in limiting cases the very simple forms:
\[I(t)=\frac{8}{3\pi}t\,,\quad t\ll 1\,,\]
\[I(t)=1-\frac{1}{4t^2},\quad t>1\,.\]

\begin{figure}[]
\resizebox{.5\textwidth}{!}{\includegraphics{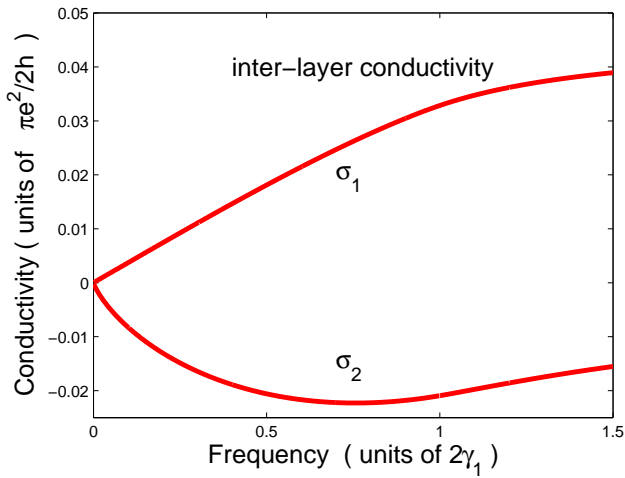}}
\caption{The real and imaginary parts of conductivity in $c-$direction;   units are the same as in  Fig. \ref{gr1}.} \label{gr3}
\end{figure}
 
  The imaginary part of the conductivity in $c-$direction is given by the $k_z-$integral 
 \begin{eqnarray}
\text{Im}~\frac{\sigma^{zz}}{\sigma_0}= \frac{4}{\pi^2}\left(\frac{\gamma_1c_0}{\hbar v}\right)^2
\int_0^{\pi/2}dz\sin^2(z)
\nonumber\\  \left[-2\frac{\gamma(z)}{\omega
}+\left(1-\frac{\gamma^2(z)}{\omega^2}\right)
 \ln{\frac{|\gamma(z)-\omega|}{\gamma(z)+\omega}}\right]\, .
 \nonumber\end{eqnarray}

The  conductivity in the $c-$direction is shown in Fig. \ref{gr3}. Compared with the in-plane conductivity, the  $c-$conductivity is less by the factor $(\gamma_1c_0/\hbar v)^2\sim 0.01$. This factor represents the squared ratio of the hopping integrals for the inter- and in-layer directions
$(\gamma_1/\gamma_0)^2\simeq \exp{(-2c_0/r_0)} $.

In conclusions, for the in-plane direction,  the optical conductivity of graphite per single atomic sheet is close to the graphene   universal conductivity.  However, the singularities, the kink in the real part and the threshold in the imaginary part, appear at the frequency $\omega=2\gamma_1$, where $\gamma_1$ is the inter-layer hopping energy for the bilayer graphene. For the $c-$direction, the conductivity is less by the parameter representing
the ratio of the inter- and in-layer hopping energies; the real part of conductivity increases linearly with the frequency and does not contain any singularities.

This work was supported by the Russian Foundation for Basic
Research (grant No. 10-02-00193-a) and by the Fondation de Cooperation Scientifique Digiteo Triangle de la Physique, 2009-069T project "BIGRAPH".

\end{document}